\documentstyle[floats,aps]{revtex}

\def\ut#1{#1\llap{\lower2ex\hbox{$\widetilde{\hphantom{#1}}$}}}

\advance\topmargin 30pt

\def\subtilde#1{#1\llap{\lower2ex\hbox{$\widetilde{\hphantom{#1}}$}}}

\begin{document} \title{Tetrads in geometrodynamics} 
\author{ Alcides Garat$^{1}$ } 

\address{1. Instituto de F\'{\i}sica, Facultad de Ciencias,
Igu\'a 4225, esq. Mataojo, Montevideo, Uruguay.} 

\date{December 7th, 2004}

\maketitle 

\begin{abstract} 
A new tetrad is introduced within the framework of geometrodynamics
for non-null electromagnetic fields.
This tetrad diagonalizes the electromagnetic stress-energy tensor and
allows for maximum simplification of the expression
of the electromagnetic field. The Einstein-Maxwell equations will also
be simplified.
\end{abstract}

\section{Introduction}
\label{introduction}

The theory of geometrodynamics \cite{MW} tried to account for the 
classical problem of the charged particle. 
It is a point of view where there is nothing except curved spacetime.
Electromagnetism is only a manifestation of the curvature. 
The electric charge was described as the flux of lines of force 
that emerges from the mouth of a small wormhole in a multiply-connected space. 
An observer with poor resolving power would see the emerging flux 
as coming from an elementary electric charge \cite{MW}.
Therefore, it was concluded that if elementary
charge can be associated with a geometrical property of the 
multiply-connected spacetime, 
there should not be a source term in Maxwell's equations. 
Geometrodynamics is not in agreement with the idea that particles and fields 
live in a geometrical background as imported entities.   
However, it presented difficulties, 
since it was not possible to find a variational principle for this theory.
If $F_{\mu\nu}$ is the electromagnetic field and 
$f_{\mu\nu}= (G^{1/2} / c^2) \: F_{\mu\nu}$ is the geometrized 
electromagnetic field, then the Einstein-Maxwell equations can be written,

\begin{eqnarray}
f^{\mu\nu}_{\:\:\:\:\:;\nu} &=& 0 \label{EM1}\\
\ast f^{\mu\nu}_{\:\:\:\:\:;\nu} &=& 0 \label{EM2}\\
R_{\mu\nu} &=& f_{\mu\lambda}\:\:f_{\nu}^{\:\:\:\lambda}
+ \ast f_{\mu\lambda}\:\ast f_{\nu}^{\:\:\:\lambda}\ , \label{EM3}
\end{eqnarray} 

where  
$\ast f_{\mu\nu}={1 \over 2}\:\epsilon_{\mu\nu\sigma\tau}\:f^{\sigma\tau}$
is the dual tensor of $f_{\mu\nu}$ (section \ref{sec:appI}). The symbol $``;''$ stands for covariant derivative with respect to the metric tensor $g_{\mu\nu}$.
The quadratic right hand side in equation (\ref{EM3}) is solved in terms
of the left hand side. Then the ``square root of the left hand side''
written in terms of the metric tensor is replaced 
in equations (\ref{EM1}-\ref{EM2}) and two sets of equations
have to be satisfied by the metric tensor.
On the one hand the Bianchi identities. They are
identically satisfied by the metric tensor, so they were not a problem.
On the other hand, a set of integrability conditions that couldn't
be derived from a variational principle.  
In this work, it is not
our goal to solve the problem of the variational principle 
in geometrodynamics.
It is our purpose to introduce a new tetrad, specifically
adapted to the geometry of electromagnetic fields in geometrodynamics.
This new tetrad could simplify the understanding of the geometry
associated with non-null electromagnetic fields.
A tetrad is a set of four linearly independent vectors 
$V_{(j)}^{\alpha}$ that could be defined at every point 
in a spacetime \cite{HS}\cite{NP}.
The index $j$ is the tetrad index and runs from one to four, naming
the vectors. It is possible then, to write at every point in a spacetime,
the components of a tensor in terms of the tetrad vectors,

\begin{eqnarray}
Z^{\alpha\beta\dots}_{\:\:\:\:\:\mu\nu\dots} &=& 
Z^{rs\dots}_{\:\:\:\:\:pq\dots}\:
V^{\alpha}_{(r)}\:V^{\beta}_{(s)}\:
V_{\mu}^{(p)}\:V_{\nu}^{(q)}\dots\ ,\label{TEXP}
\end{eqnarray}

where the quantities $Z^{rs\dots}_{\:\:\:\:\:pq\dots}$ are the tetrad 
components of the tensor. Our purpose is to find a tetrad
in geometrodynamics that diagonalizes the stress-energy tensor
and simplifies the expression of the electromagnetic field.
We need to understand first the concept of duality rotations.
A duality rotation of the electromagnetic field is defined as \cite{MW},

\begin{equation}
e^{\ast \alpha} f_{\mu\nu} = f_{\mu\nu} \: \cos\alpha +
\ast f_{\mu\nu} \: \sin\alpha\ ,\label{drintro}
\end{equation}

where $\alpha$ is a scalar called the complexion.
At every point in spacetime there is a duality rotation by an angle
$-\alpha$ that transforms a non-null electromagnetic field into 
an extremal field,

\begin{equation}
\xi_{\mu\nu} = e^{-\ast \alpha} f_{\mu\nu}\ .\label{dref} 
\end{equation}

Extremal fields are essentially electric fields and they satisfy,

\begin{equation}
\xi_{\mu\nu} \ast \xi^{\mu\nu}= 0\ . 
\end{equation}

The tetrad that diagonalizes the stress-energy tensor will
be written in terms of the extremal field $\xi_{\mu\nu}$,
and two other vector fields $X^{\alpha}$ and $Y^{\alpha}$. 
It will be proved that these two extra vector
fields are available freedom that we have in the construction of a 
general tetrad for non-null electromagnetic fields.
The fact that in geometrodynamics Maxwell's equations 
(\ref{EM1}-\ref{EM2}) have
zero source terms, introduces the existence of two potential vector fields,
$A^{\alpha}$ and $\ast A^{\alpha}$, natural candidates 
for a particular and explicit choice or example 
of $X^{\alpha}$ and $Y^{\alpha}$. 
In this particular example, an unexpected question will arise at this point.
If our tetrad involves in its construction the potential vectors
$A^{\alpha}$ and $\ast A^{\alpha}$, how is the tetrad going to be affected
by electromagnetic gauge transformations. 
The geometry of electromagnetic fields defines at every point in 
spacetime two planes related to the symmetries of the stress-energy tensor
\cite{SCH}.
Gauge transformations 
$A^{\alpha} \rightarrow A^{\alpha} + \Lambda^{,\alpha}$,
with $\Lambda$ a scalar function,
that leave invariant the electromagnetic field,
will generate proper and improper Lorentz transformations on
one of the planes.
Gauge transformations 
$\ast A^{\alpha} \rightarrow \ast A^{\alpha} + \ast \Lambda^{,\alpha}$,
with $\ast \Lambda$ a scalar function,
that leave invariant the dual of the electromagnetic
field will generate spatial rotations on the other plane.
The possibility of introducing null tetrads in geometrodynamics will
also be explored. Finally, the general tetrad will be studied for 
non-null electromagnetic fields where $X^{\alpha}$ and $Y^{\alpha}$
are considered as generic fields,
and the Einstein-Maxwell equations
in this general tetrad will be discussed. The explicit example introduced
and all its properties will be useful in understanding the general case.  
The remainder of this paper will be organized as follows.
The tetrad that diagonalizes the stress-energy tensor will be introduced in
section \ref{diagonal}. The available freedom that we have in building
this tetrad will be analyzed in section \ref{potentials}. The geometrical
implications of gauge transformations will be discussed 
in \ref{gaugegeometry}. An isomorphism between the local gauge group and local Lorentz transformations on blades one and two will be included in \ref{Groupiso}. 
The normalized tetrad will be studied in \ref{tetrads}. In this last
section a new null tetrad will also be introduced.
The general case and the Einstein-Maxwell equations written
in terms of the general tetrad will be discussed in \ref{Gentetrads}.
Throughout the paper we
use the conventions of \cite{MW}. In particular we
use a metric with sign conventions -+++. The only difference in notation
with \cite{MW} will be that
we will call our geometrized electromagnetic potential $A^{\alpha}$,
where $f_{\mu\nu}=A_{\nu ;\mu} - A_{\mu ;\nu}$ is the geometrized 
electromagnetic field $f_{\mu\nu}= (G^{1/2} / c^2) \: F_{\mu\nu}$.

\section{Diagonalization of the stress-energy tensor}
\label{diagonal}

The stress-energy tensor according to equation (14a)
in \cite{MW}, can be written as,

\begin{equation}
T_{\mu\nu}= f_{\mu\lambda}\:\:f_{\nu}^{\:\:\:\lambda}
+ \ast f_{\mu\lambda}\:\ast f_{\nu}^{\:\:\:\lambda}\ ,\label{TEM}
\end{equation}

where 
$\ast f_{\mu\nu}={1 \over 2}\:\epsilon_{\mu\nu\sigma\tau}\:f^{\sigma\tau}$
is the dual tensor of $f_{\mu\nu}$. The tensor 
$\epsilon_{\mu\nu\sigma\tau}$ is studied in appendix \ref{sec:appI}.

The duality rotation given by equation (59) in\cite{MW},

\begin{equation}
f_{\mu\nu} = \xi_{\mu\nu} \: \cos\alpha +
\ast\xi_{\mu\nu} \: \sin\alpha\ ,\label{dr}
\end{equation}

allows us to express the stress-energy tensor in terms of the extremal field,

\begin{equation}
T_{\mu\nu}=\xi_{\mu\lambda}\:\:\xi_{\nu}^{\:\:\:\lambda}
+ \ast \xi_{\mu\lambda}\:\ast \xi_{\nu}^{\:\:\:\lambda}\ .\label{TEMDR}
\end{equation}

The extremal field $\xi_{\mu\nu}$ and the scalar complexion $\alpha$ are 
defined through equations (22-25) in \cite{MW}. 
It is our purpose to find a tetrad in which the stress-energy tensor
is diagonal. This tetrad would simplify the analysis of the geometrical
properties of the electromagnetic field. 
There are four tetrad vectors that at every point in spacetime 
diagonalize the stress-energy tensor in geometrodynamics,

\begin{eqnarray}
V_{(1)}^{\alpha} &=& \xi^{\alpha\lambda}\:\xi_{\rho\lambda}\:X^{\rho}
\label{V1}\\
V_{(2)}^{\alpha} &=& \sqrt{-Q/2} \: \xi^{\alpha\lambda} \: X_{\lambda}
\label{V2}\\
V_{(3)}^{\alpha} &=& \sqrt{-Q/2} \: \ast \xi^{\alpha\lambda} \: Y_{\lambda}
\label{V3}\\
V_{(4)}^{\alpha} &=& \ast \xi^{\alpha\lambda}\: \ast \xi_{\rho\lambda}
\:Y^{\rho}\ ,\label{V4}
\end{eqnarray}

where $Q=\xi_{\mu\nu}\:\xi^{\mu\nu}=-\sqrt{T_{\mu\nu}T^{\mu\nu}}$ 
according to equations (39) in \cite{MW}. $Q$ is assumed not to be zero,
because we are dealing with non-null electromagnetic fields.
We are free to choose the vector fields $X^{\alpha}$ and $Y^{\alpha}$, as
long as the four vector fields (\ref{V1}-\ref{V4}) are not trivial.
Two identities in the extremal field are going to be used extensively
in this work, in particular, to prove that tetrad (\ref{V1}-\ref{V4}) 
diagonalizes the stress-energy tensor. The first identity is given by 
equation (64) in \cite{MW},

\begin{eqnarray}
\xi_{\alpha\mu}\:\ast \xi^{\mu\nu} &=& 0\ .\label{i1}
\end{eqnarray} 

In order to find the second identity we need equation (15) in \cite{MW},

\begin{eqnarray}
f_{\mu\alpha}\:f^{\nu\alpha} - 
\ast f_{\mu\alpha}\: \ast f^{\nu\alpha} &=& \frac{1}{2} \:\: 
\delta_{\mu}^{\:\:\:\nu}\ f_{\alpha\beta}f^{\alpha\beta}\ . \label{if}
\end{eqnarray} 

When we replace (\ref{dr}) in (\ref{if}) and make use of 
(\ref{i1}), the second identity is found,

\begin{eqnarray}
\xi_{\mu\alpha}\:\xi^{\nu\alpha} - 
\ast \xi_{\mu\alpha}\: \ast \xi^{\nu\alpha} &=& \frac{1}{2} 
\: \delta_{\mu}^{\:\:\:\nu}\ Q \ .\label{i2}
\end{eqnarray}

When we make iterative use of (\ref{i1}) and (\ref{i2}) we find,

\begin{eqnarray}
V_{(1)}^{\alpha}\:T_{\alpha}^{\:\:\:\beta} &=& \frac{Q}{2}\:V_{(1)}^{\beta}
\label{EV1}\\
V_{(2)}^{\alpha}\:T_{\alpha}^{\:\:\:\beta} &=& \frac{Q}{2}\:V_{(2)}^{\beta}
\label{EV2}\\
V_{(3)}^{\alpha}\:T_{\alpha}^{\:\:\:\beta} &=& -\frac{Q}{2}\:V_{(3)}^{\beta}
\label{EV3}\\
V_{(4)}^{\alpha}\:T_{\alpha}^{\:\:\:\beta} &=& -\frac{Q}{2}\:V_{(4)}^{\beta}\ .
\label{EV4}
\end{eqnarray} 

In \cite{MW} the stress-energy tensor was diagonalized through the use
of a Minkowskian frame in which the equation for this tensor 
was given in (34) and (38). 
In this work, we give the explicit expression for the
tetrad in which the stress-energy tensor is diagonal. 
The freedom we have to choose the vector fields $X^{\alpha}$ and $Y^{\alpha}$,
represents available freedom that we have to choose the tetrad.  
If we make use of equations (\ref{i1}) and (\ref{i2}),
it is straightforward to prove that (\ref{V1}-\ref{V4}) 
is a set of orthogonal vectors.

\section{Electromagnetic potentials in geometrodynamics}
\label{potentials}

Our goal is to simplify as much as we can the expression of the
electromagnetic field through the use of an orthonormal tetrad,
so its geometrical properties 
can be understood in an easier way. As it was mentioned above we would
like to show this simplification through an explicit example by
making a convenient and particular choice of the vector fields 
$X^{\alpha}$ and $Y^{\alpha}$.
In geometrodynamics, the Maxwell equations,

\begin{eqnarray}
f^{\mu\nu}_{\:\:\:\:\:;\nu} &=& 0 \label{L1}\nonumber\\
\ast f^{\mu\nu}_{\:\:\:\:\:;\nu} &=& 0 \ , \label{L2}
\end{eqnarray}

are telling us that two potential vector fields exist,

\begin{eqnarray}
f_{\mu\nu} &=& A_{\nu ;\mu} - A_{\mu ;\nu}\label{ER}\nonumber\\
\ast f_{\mu\nu} &=& \ast A_{\nu ;\mu} - \ast A_{\mu ;\nu} \ .\label{DER}
\end{eqnarray} 

For instance, in the Reissner-Nordstrom geometry the only non-zero
electromagnetic tensor component is $f_{tr}=A_{r;t} - A_{t;r}$ and its dual
$\ast f_{\theta\phi}=\ast A_{\phi;\theta} - \ast A_{\theta;\phi}$.
The vector fields $A^{\alpha}$ and $\ast A^{\alpha}$ represent a 
possible choice in geometrodynamics for the vectors
$X^{\alpha}$ and $Y^{\alpha}$. It is not meant that the two vector
fields have independence from each other, it is just a convenient choice
for a particular example.
This choice allows us to write the new tetrad as,

\begin{eqnarray}
V_{(1)}^{\alpha} &=& \xi^{\alpha\lambda}\:\xi_{\rho\lambda}\:A^{\rho}
\label{V1A}\\
V_{(2)}^{\alpha} &=& \sqrt{-Q/2} \: \xi^{\alpha\lambda} \: A_{\lambda}
\label{V2A}\\
V_{(3)}^{\alpha} &=& \sqrt{-Q/2} \: \ast \xi^{\alpha\lambda} 
\: \ast A_{\lambda}\label{V3A}\\
V_{(4)}^{\alpha} &=& \ast \xi^{\alpha\lambda}\: \ast \xi_{\rho\lambda}
\:\ast A^{\rho}\ .\label{V4A}
\end{eqnarray}

A further justification for the choice $X^{\alpha}=A^{\alpha}$ and 
$Y^{\alpha}=\ast A^{\alpha}$ could be illustrated through the 
Reissner-Nordstrom geometry. In this particular geometry, $f_{tr}=\xi_{tr}$
and $\ast f_{\theta\phi}=\ast \xi_{\theta\phi}$, therefore, $A_{\theta}=0$
and $A_{\phi}=0$. Then, for the last two tetrad vectors (\ref{V3A}-\ref{V4A}), 
the choice $Y^{\alpha}=\ast A^{\alpha}$ becomes meaningful under the light of 
this particular extreme case, when basically there is no magnetic field.
However, we have to be careful about the choice we made for 
$X^{\alpha}$ and $Y^{\alpha}$. The normalization of the tetrad vectors
(\ref{V1A}-\ref{V4A}) requires to know the values of the invariant
quantities,

\begin{eqnarray}
V_{(1)}^{\alpha}\:V_{(1)\alpha} &=& 
(Q/2) \: A_{\mu} \ \xi^{\mu\sigma} \ 
\xi_{\nu\sigma} \ A^{\nu} \label{S1}\\
V_{(2)}^{\alpha}\:V_{(2)\alpha} &=& 
(-Q/2) \: A_{\mu} \ \xi^{\mu\sigma} \ 
\xi_{\nu\sigma} \ A^{\nu} \label{S2}\\
V_{(3)}^{\alpha}\:V_{(3)\alpha} &=& 
(-Q/2) \: \ast A_{\mu} \ \ast \xi^{\mu\sigma} \ 
\ast \xi_{\nu\sigma} \ \ast A^{\nu} \label{S3}\\
V_{(4)}^{\alpha}\:V_{(4)\alpha} &=& 
(-Q/2) \: \ast A_{\mu} \ \ast \xi^{\mu\sigma} \ 
\ast \xi_{\nu\sigma} \ \ast A^{\nu}\ . \label{S4}
\end{eqnarray}

Then, it is convenient to calculate the invariants 
(\ref{S1}-\ref{S4}) in the Minkowski 
reference frame given by equations (38-39) in \cite{MW},

\begin{eqnarray}
V_{(1)}^{\alpha}\:V_{(1)\alpha} &=& 
(\xi_{01})^4 \: (A^{0}\:A_{0} + A^{1}\:A_{1}) \label{SM1}\\
V_{(2)}^{\alpha}\:V_{(2)\alpha} &=& 
-(\xi_{01})^4 \: (A^{0}\:A_{0} + A^{1}\:A_{1}) \label{SM2}\\
V_{(3)}^{\alpha}\:V_{(3)\alpha} &=& 
(\xi_{01})^4 \: (\ast A^{2} \: \ast A_{2} + 
\ast A^{3} \: \ast A_{3}) \label{SM3}\\
V_{(4)}^{\alpha}\:V_{(4)\alpha} &=& 
(\xi_{01})^4 \: (\ast A^{2} \: \ast A_{2} + 
\ast A^{3} \: \ast A_{3})\ . \label{SM4}
\end{eqnarray}

Several cases arise. Since $\xi_{01} \neq 0$ because the electromagnetic field
is non-null ($Q \neq 0$), the two vector components
$A_{0}$ and $A_{1}$ cannot be simultaneously zero. 
If in a region of spacetime or in the
whole spacetime $A^{0}\:A_{0} + A^{1}\:A_{1} > 0$, 
it would be necessary to switch the first two tetrad vectors,
$V_{(1)}^{\alpha} = \sqrt{-Q/2} \: \xi^{\alpha\lambda} \: A_{\lambda}$
and $V_{(2)}^{\alpha} = \xi^{\alpha\lambda}\:\xi_{\rho\lambda}\:A^{\rho}$. 
If  $A^{0}\:A_{0} + A^{1}\:A_{1} = 0$, then we have 
$V_{(1)}^{\alpha}\:T_{\alpha}^{\:\:\:\beta}\:V_{(1)\beta}=0$ and 
$V_{(2)}^{\alpha}\:T_{\alpha}^{\:\:\:\beta}\:V_{(2)\beta}=0$. We are not
dealing with this kind of situation in this work. Another
choice of $X^{\alpha}$ would be necessary at the points
where $A^{0}\:A_{0} + A^{1}\:A_{1} = 0$.
It is clear that if $A^{0}\:A_{0} + A^{1}\:A_{1} < 0$, the 
following equations hold,

\begin{eqnarray}
-V_{(1)}^{\alpha}\:V_{(1)\alpha} &=& V_{(2)}^{\alpha}\:V_{(2)\alpha} > 0 \\
V_{(3)}^{\alpha}\:V_{(3)\alpha} &=& V_{(4)}^{\alpha}\:V_{(4)\alpha} > 0 \ .
\label{TS12}
\end{eqnarray}

The treatment of the problem of building tetrad vectors in geometrodynamics,
studying their gauge transformation properties and building null tetrads
would be analogous for the two cases when $A^{0}\:A_{0} + A^{1}\:A_{1} > 0$ 
or $A^{0}\:A_{0} + A^{1}\:A_{1} < 0$. 
The treatment of just one of the these two possible cases,
would automatically provide the framework and the ideas to solve the other one.
That is why we choose to analyze in detail $A^{0}\:A_{0} + A^{1}\:A_{1} < 0$.


\section{gauge geometry}
\label{gaugegeometry}

Once we make the choice $X^{\alpha}=A^{\alpha}$ and 
$Y^{\alpha}=\ast A^{\alpha}$ the question about the geometrical
implications of electromagnetic gauge transformations arises.
When we make the transformation,

\begin{eqnarray}
A_{\alpha} \rightarrow A_{\alpha} + \Lambda_{,\alpha}\ , \label{G1}
\end{eqnarray}

$f_{\mu\nu}$ remains invariant, and the transformation,

\begin{eqnarray}
\ast A_{\alpha} \rightarrow \ast A_{\alpha} + 
\ast \Lambda_{,\alpha}\ , \label{G2}
\end{eqnarray}

leaves $\ast f_{\mu\nu}$ invariant,
as long as the functions $\Lambda$ and $\ast \Lambda$ are
scalars. It is valid to ask how the tetrad vectors (\ref{V1A}-\ref{V2A}) 
are going to transform under (\ref{G1}) and (\ref{V3A}-\ref{V4A}) under
(\ref{G2}). Schouten defined what he called, a two-bladed structure
in a spacetime \cite{SCH}. 
These blades are the planes determined by the pairs 
($V_{(1)}^{\alpha}, V_{(2)}^{\alpha}$) and 
($V_{(3)}^{\alpha}, V_{(4)}^{\alpha}$).
For simplicity we will study first what we call, gauge transformations
associated with blade one, the blade generated by the pair 
($V_{(1)}^{\alpha}, V_{(2)}^{\alpha}$). Later we will study similar 
transformations on blade two, the blade generated
by ($V_{(3)}^{\alpha}, V_{(4)}^{\alpha}$).

\subsection{Gauge transformations on blade one}
\label{gaugegeometry1}

In order to simplify the notation we are going to write 
$\Lambda_{,\alpha}=\Lambda_{\alpha}$.
First we study the change in (\ref{V1A}-\ref{V2A}) under (\ref{G1}),

\begin{eqnarray}
\tilde{V}_{(1)}^{\alpha} &=& V_{(1)}^{\alpha} + 
\xi^{\alpha\lambda}\:\xi_{\rho\lambda}\:\Lambda^{\rho}\label{TU}\\
\tilde{V}_{(2)}^{\alpha} &=& V_{(2)}^{\alpha} + 
\sqrt{-Q/2}\:\xi^{\alpha\lambda}\:\Lambda_{\lambda}\ ,\label{TV}
\end{eqnarray}

The second term on the right hand side of (\ref{TU}) has the 
orthogonality properties,

\begin{eqnarray} 
\xi^{\alpha\lambda}\:\xi_{\rho\lambda}\:\Lambda^{\rho}\:V_{(3)\alpha}=
\xi^{\alpha\lambda}\:\xi_{\rho\lambda}\:\Lambda^{\rho}\:V_{(4)\alpha}=0
\ .\label{OTU}
\end{eqnarray}

The second term on the right hand side of (\ref{TV}) has 
similar orthogonality properties,

\begin{eqnarray} 
\sqrt{-Q/2}\:\xi^{\alpha\lambda}\:\Lambda_{\lambda}\:V_{(3)\alpha}=
\sqrt{-Q/2}\:\xi^{\alpha\lambda}\:\Lambda_{\lambda}\:V_{(4)\alpha}=0
\ .\label{OTV}
\end{eqnarray}

Since the four vectors (\ref{V1A}-\ref{V4A}) are independent and orthogonal,
equations (\ref{OTU}-\ref{OTV}) imply that the second terms on the 
right hand sides of (\ref{TU}-\ref{TV}) must be a linear combination of 
the vectors (\ref{V1A}-\ref{V2A}). We proceed then to write equations
(\ref{TU}-\ref{TV}) as,

\begin{eqnarray} 
\tilde{V}_{(1)}^{\alpha} &=& V_{(1)}^{\alpha} + 
C\:V_{(1)}^{\alpha} + D\:V_{(2)}^{\alpha}\label{TUN}\\
\tilde{V}_{(2)}^{\alpha} &=& V_{(2)}^{\alpha} + 
E\:V_{(1)}^{\alpha} + F\:V_{(2)}^{\alpha}\ .\label{TVN}
\end{eqnarray}

From equations (\ref{TU}) and (\ref{TUN}) we know that,

\begin{equation}
\xi^{\alpha\lambda}\:\xi_{\rho\lambda}\:\Lambda^{\rho}=
C\:V_{(1)}^{\alpha} + D\:V_{(2)}^{\alpha}\ ,\label{TUNEXP}
\end{equation}

and from equations (\ref{TV}) and (\ref{TVN}) we have,

\begin{equation}
\sqrt{-Q/2}\:\xi^{\alpha\lambda}\:\Lambda_{\lambda}=
E\:V_{(1)}^{\alpha} + F\:V_{(2)}^{\alpha}\ .\label{TVNEXP}
\end{equation}

Making use of identities (\ref{i1}-\ref{i2}) it can be proved that,

\begin{eqnarray} 
\xi_{\alpha\sigma}\:V_{(1)}^{\alpha}&=&
\sqrt{-Q/2}\:V_{(2)\sigma}\label{OR1}\\
\xi_{\alpha\sigma}\:V_{(2)}^{\alpha}&=&
\sqrt{-Q/2}\:V_{(1)\sigma}\ .\label{OR2}
\end{eqnarray}

If we contract equation (\ref{TUNEXP}) with $\xi_{\alpha\sigma}$ and
make use of identities (\ref{i1}-\ref{i2}) and equations 
(\ref{OR1}-\ref{OR2}) we get,

\begin{equation}
\sqrt{-Q/2}\:\xi_{\sigma\lambda}\:\Lambda^{\lambda}=
C\:V_{(2)\sigma} + D\:V_{(1)\sigma}\ .\label{BP}
\end{equation}

This last equation means that,

\begin{equation}
\tilde{V}_{(2)}^{\alpha} = V_{(2)}^{\alpha} + 
C\:V_{(2)}^{\alpha} + D\:V_{(1)}^{\alpha}\ .\label{TVB}
\end{equation}

Then, from (\ref{TVN}) and (\ref{TVB}) we have the following 
relations between coefficients, 

\begin{eqnarray}
E &=& D \\
F &=& C \ .
\end{eqnarray}

Contracting equation (\ref{BP}) with $V_{(1)}^{\sigma}$ and
$V_{(2)}^{\sigma}$ it can be found that,

\begin{eqnarray} 
C&=&(-Q/2)\:V_{(1)\sigma}\:\Lambda^{\sigma} / (\:V_{(2)\beta}\:
V_{(2)}^{\beta}\:)\label{COEFFC}\\
D&=&(-Q/2)\:V_{(2)\sigma}\:\Lambda^{\sigma} / (\:V_{(1)\beta}\:
V_{(1)}^{\beta}\:)\ .\label{COEFFD}
\end{eqnarray}

We would like to calculate the norm of the transformed vectors
$\tilde{V}_{(1)}^{\alpha}$ and $\tilde{V}_{(2)}^{\alpha}$,

\begin{eqnarray}
\tilde{V}_{(1)}^{\alpha}\:\tilde{V}_{(1)\alpha} &=& 
[(1+C)^2-D^2]\:V_{(1)}^{\alpha}\:V_{(1)\alpha}\label{FP}\\
\tilde{V}_{(2)}^{\alpha}\:\tilde{V}_{(2)\alpha} &=& 
[(1+C)^2-D^2]\:V_{(2)}^{\alpha}\:V_{(2)\alpha}\ ,\label{SP}
\end{eqnarray}
 
where the relation $V_{(1)}^{\alpha}\:V_{(1)\alpha}=
-V_{(2)}^{\alpha}\:V_{(2)\alpha}$ has been used.
In order for these transformations to keep the timelike or spacelike
character of $V_{(1)}^{\alpha}$ and $V_{(2)}^{\alpha}$ the condition
$[(1+C)^2-D^2]>0$ must be satisfied. If this condition is
fulfilled, then we can normalize the transformed vectors
$\tilde{V}_{(1)}^{\alpha}$ and $\tilde{V}_{(2)}^{\alpha}$ as follows,

\begin{eqnarray}
{\tilde{V}_{(1)}^{\alpha} 
\over \sqrt{-\tilde{V}_{(1)}^{\beta}\:\tilde{V}_{(1)\beta}}}&=& 
{(1+C) \over \sqrt{(1+C)^2-D^2}}
\:{V_{(1)}^{\alpha} \over \sqrt{-V_{(1)}^{\beta}\:V_{(1)\beta}}}+
{D \over \sqrt{(1+C)^2-D^2}}
\:{V_{(2)}^{\alpha} \over \sqrt{V_{(2)}^{\beta}\:V_{(2)\beta}}}\label{TN1}\\
{\tilde{V}_{(2)}^{\alpha} 
\over \sqrt{\tilde{V}_{(2)}^{\beta}\:\tilde{V}_{(2)\beta}}}&=& 
{D \over \sqrt{(1+C)^2-D^2}}
\:{V_{(1)}^{\alpha} \over \sqrt{-V_{(1)}^{\beta}\:V_{(1)\beta}}} +
{(1+C) \over \sqrt{(1+C)^2-D^2}}
\:{V_{(2)}^{\alpha} \over \sqrt{V_{(2)}^{\beta}\:V_{(2)\beta}}}\ .
\label{TN2}
\end{eqnarray}

The condition $[(1+C)^2-D^2]>0$ allows for two possible situations,
$1+C > 0$ or $1+C < 0$.
For the particular case when $1+C > 0$, the transformations 
(\ref{TN1}-\ref{TN2}) are telling us that an
electromagnetic gauge transformation on the vector field $A^{\alpha}$,
that leaves invariant the electromagnetic field $f_{\mu\nu}$, generates
a boost transformation on the normalized tetrad vector fields 
$\left({V_{(1)}^{\alpha} \over \sqrt{-V_{(1)}^{\beta}\:V_{(1)\beta}}},
{V_{(2)}^{\alpha} \over \sqrt{V_{(2)}^{\beta}\:V_{(2)\beta}}}\right)$. 
For the case $1+C < 0$, equations (\ref{TN1}-\ref{TN2}) can be rewritten,

\begin{eqnarray}
{\tilde{V}_{(1)}^{\alpha} 
\over \sqrt{-\tilde{V}_{(1)}^{\beta}\:\tilde{V}_{(1)\beta}}}&=& 
{[-(1+C)] \over \sqrt{(1+C)^2-D^2}}
\:{\left(-V_{(1)}^{\alpha}\right) 
\over \sqrt{-V_{(1)}^{\beta}\:V_{(1)\beta}}}+
{[-D] \over \sqrt{(1+C)^2-D^2}}
\:{\left(-V_{(2)}^{\alpha}\right) \over \sqrt{V_{(2)}^{\beta}
\:V_{(2)\beta}}}\label{TN1SC}\\
{\tilde{V}_{(2)}^{\alpha} 
\over \sqrt{\tilde{V}_{(2)}^{\beta}\:\tilde{V}_{(2)\beta}}}&=& 
{[-D] \over \sqrt{(1+C)^2-D^2}}
\:{\left(-V_{(1)}^{\alpha}\right) 
\over \sqrt{-V_{(1)}^{\beta}\:V_{(1)\beta}}} +
{[-(1+C)] \over \sqrt{(1+C)^2-D^2}}
\:{\left(-V_{(2)}^{\alpha}\right) \over \sqrt{V_{(2)}^{\beta}
\:V_{(2)\beta}}}\ . \label{TN2SC}
\end{eqnarray}

Equations (\ref{TN1SC}-\ref{TN2SC}) represent the composition of two
transformations. An inversion of the
normalized tetrad vector fields 
$\left({V_{(1)}^{\alpha} \over \sqrt{-V_{(1)}^{\beta}\:V_{(1)\beta}}},
{V_{(2)}^{\alpha} \over \sqrt{V_{(2)}^{\beta}\:V_{(2)\beta}}}\right)$,
and a boost. 
If the case is that $[(1+C)^2-D^2]<0$, the vectors $V_{(1)}^{\alpha}$ and $V_{(2)}^{\alpha}$ are going to change their timelike or spacelike character,

\begin{eqnarray}
\tilde{V}_{(1)}^{\alpha}\:\tilde{V}_{(1)\alpha} &=& 
[-(1+C)^2+D^2]\:(-V_{(1)}^{\alpha}\:V_{(1)\alpha})\label{FPI}\\
(-\tilde{V}_{(2)}^{\alpha}\:\tilde{V}_{(2)\alpha}) &=& 
[-(1+C)^2+D^2]\:V_{(2)}^{\alpha}\:V_{(2)\alpha}\ .\label{SPI}
\end{eqnarray}
 
These are improper transformations on blade one. The normalized tetrad vectors $V_{(1)}^{\alpha}$ and $V_{(2)}^{\alpha}$ transform as,

\begin{eqnarray}
{\tilde{V}_{(1)}^{\alpha} 
\over \sqrt{\tilde{V}_{(1)}^{\beta}\:\tilde{V}_{(1)\beta}}}&=& 
{(1+C) \over \sqrt{-(1+C)^2+D^2}}
\:{V_{(1)}^{\alpha} \over \sqrt{-V_{(1)}^{\beta}\:V_{(1)\beta}}}+
{D \over \sqrt{-(1+C)^2+D^2}}
\:{V_{(2)}^{\alpha} \over \sqrt{V_{(2)}^{\beta}\:V_{(2)\beta}}}\label{TN1I}\\
{\tilde{V}_{(2)}^{\alpha} 
\over \sqrt{-\tilde{V}_{(2)}^{\beta}\:\tilde{V}_{(2)\beta}}}&=& 
{D \over \sqrt{-(1+C)^2+D^2}}
\:{V_{(1)}^{\alpha} \over \sqrt{-V_{(1)}^{\beta}\:V_{(1)\beta}}} +
{(1+C) \over \sqrt{-(1+C)^2+D^2}}
\:{V_{(2)}^{\alpha} \over \sqrt{V_{(2)}^{\beta}\:V_{(2)\beta}}}\ .
\label{TN2I}
\end{eqnarray}

For $D > 0$ and $1+C > 0$ these transformations (\ref{TN1I}-\ref{TN2I}) represent improper space inversions on blade one. If $D > 0$ and $1+C < 0$, equations (\ref{TN1I}-\ref{TN2I}) are improper time reversal transformations on blade one \cite{WE}. If the case is that $D < 0$, we can proceed to analyze in analogy to (\ref{TN1SC}-\ref{TN2SC}). Then, the normalized tetrad vectors transform as,

\begin{eqnarray}
{\tilde{V}_{(1)}^{\alpha} 
\over \sqrt{\tilde{V}_{(1)}^{\beta}\:\tilde{V}_{(1)\beta}}}&=& 
{[-(1+C)] \over \sqrt{-(1+C)^2+D^2}}
\:{\left(-V_{(1)}^{\alpha}\right) 
\over \sqrt{-V_{(1)}^{\beta}\:V_{(1)\beta}}}+
{[-D] \over \sqrt{-(1+C)^2+D^2}}
\:{\left(-V_{(2)}^{\alpha}\right) \over \sqrt{V_{(2)}^{\beta}
\:V_{(2)\beta}}}\label{TN1SCI}\\
{\tilde{V}_{(2)}^{\alpha} 
\over \sqrt{-\tilde{V}_{(2)}^{\beta}\:\tilde{V}_{(2)\beta}}}&=& 
{[-D] \over \sqrt{-(1+C)^2+D^2}}
\:{\left(-V_{(1)}^{\alpha}\right) 
\over \sqrt{-V_{(1)}^{\beta}\:V_{(1)\beta}}} +
{[-(1+C)] \over \sqrt{-(1+C)^2+D^2}}
\:{\left(-V_{(2)}^{\alpha}\right) \over \sqrt{V_{(2)}^{\beta}
\:V_{(2)\beta}}}\ . \label{TN2SCI}
\end{eqnarray}

For $D < 0$ and $1+C < 0$ these transformations (\ref{TN1SCI}-\ref{TN2SCI}) represent the composition of inversions, and improper space inversions on blade one. If $D < 0$ and $1+C > 0$, equations (\ref{TN1SCI}-\ref{TN2SCI}) are inversions composed with improper time reversal transformations on blade one \cite{WE}. For $D = 1+C$ we can see using equations (\ref{TUN}) and (\ref{TVB}) that,

\begin{eqnarray}
\tilde{V}_{(1)}^{\alpha} &=& (1+C)\:V_{(1)}^{\alpha} + 
(1+C)\:V_{(2)}^{\alpha}\label{TUNNULL}\\
\tilde{V}_{(2)}^{\alpha} &=& (1+C)\:V_{(2)}^{\alpha} + 
(1+C)\:V_{(1)}^{\alpha}\ .\label{TVBNULL}
\end{eqnarray}

Equations (\ref{TUNNULL}-\ref{TVBNULL}) show that any vector on blade one transforms as,

\begin{equation}
A\: V_{(1)}^{\alpha} + B\: V_{(2)}^{\alpha} \rightarrow 
A\: \tilde{V}_{(1)}^{\alpha} + B\: \tilde{V}_{(2)}^{\alpha} = (1+C)\:(A+B)\:
( V_{(1)}^{\alpha} +  V_{(2)}^{\alpha})\ .\label{NONINJ}
\end{equation}

This is clearly a non-injective transformation. At the same time we know that there is an inverse transformation,

\begin{eqnarray}
\tilde{\tilde{V}}_{(1)}^{\alpha} &=& \tilde{V}_{(1)}^{\alpha} - 
\xi^{\alpha\lambda}\:\xi_{\rho\lambda}\:\Lambda^{\rho} = V_{(1)}^{\alpha}\label{ITUNULL}\\
\tilde{\tilde{V}}_{(2)}^{\alpha} &=& \tilde{V}_{(2)}^{\alpha} - 
\sqrt{-Q/2}\:\xi^{\alpha\lambda}\:\Lambda_{\lambda} = V_{(2)}^{\alpha}\ .\label{ITVNULL}
\end{eqnarray}

Then, the conclusion must be, that there could not exist a scalar function that satisfies the initial assumption $D = 1+C$. Analogous for $D = -(1+C)$.

\subsection{Gauge transformations on blade two}
\label{gaugegeometry2}

The change in notation  
$\ast \Lambda_{,\alpha}=\ast \Lambda_{\alpha}$ is going to be adopted.
In this section we are interested in the study of the change in 
(\ref{V3A}-\ref{V4A}) under (\ref{G2}),

\begin{eqnarray}
\tilde{V}_{(3)}^{\alpha} &=& V_{(3)}^{\alpha} + 
\sqrt{-Q/2}\:\ast \xi^{\alpha\lambda}\:\ast \Lambda_{\lambda}\label{TZ}\\
\tilde{V}_{(4)}^{\alpha} &=& V_{(4)}^{\alpha} + 
\ast \xi^{\alpha\lambda}\:\ast \xi_{\rho\lambda}\:\ast \Lambda^{\rho}
\ .\label{TW}
\end{eqnarray}

The second term on the right hand side of (\ref{TZ}) has the 
orthogonality properties,

\begin{eqnarray} 
\sqrt{-Q/2}\:\ast \xi^{\alpha\lambda}\:\ast \Lambda_{\lambda}\:V_{(1)\alpha}=
\sqrt{-Q/2}\:\ast \xi^{\alpha\lambda}\:\ast \Lambda_{\lambda}\:V_{(2)\alpha}=0
\ .\label{OTZ}
\end{eqnarray}

The second term on the right hand side of (\ref{TW}) has 
similar orthogonality properties,

\begin{eqnarray} 
\ast \xi^{\alpha\lambda}\:\ast \xi_{\rho\lambda}
\:\ast \Lambda^{\rho}\:V_{(1)\alpha}=
\ast \xi^{\alpha\lambda}\:\ast \xi_{\rho\lambda}
\:\ast \Lambda^{\rho}\:V_{(2)\alpha}=0
\ .\label{OTW}
\end{eqnarray}

Since the four vectors (\ref{V1A}-\ref{V4A}) are independent and orthogonal,
equations (\ref{OTZ}-\ref{OTW}) imply that the second terms on the 
right hand sides of (\ref{TZ}-\ref{TW}) must be a linear combination of 
the vectors (\ref{V3A}-\ref{V4A}). We proceed then to write equations
(\ref{TZ}-\ref{TW}) as,

\begin{eqnarray} 
\tilde{V}_{(3)}^{\alpha} &=& V_{(3)}^{\alpha} + 
K\:V_{(3)}^{\alpha} + L\:V_{(4)}^{\alpha}\label{TZN}\\
\tilde{V}_{(4)}^{\alpha} &=& V_{(4)}^{\alpha} + 
M\:V_{(3)}^{\alpha} + N\:V_{(4)}^{\alpha}\ .\label{TWN}
\end{eqnarray}

From equations (\ref{TZ}) and (\ref{TZN}) we know that,

\begin{equation}
\sqrt{-Q/2}\:\ast \xi^{\alpha\lambda}\:\ast \Lambda_{\lambda}=
K\:V_{(3)}^{\alpha} + L\:V_{(4)}^{\alpha}\ .\label{TZNEXP}
\end{equation}

and from equations (\ref{TW}) and (\ref{TWN}) we have,

\begin{equation}
\ast \xi^{\alpha\lambda}\:\ast \xi_{\rho\lambda}\:\ast \Lambda^{\rho}=
M\:V_{(3)}^{\alpha} + N\:V_{(4)}^{\alpha}\ .\label{TWNEXP}
\end{equation}

Making use of identities (\ref{i1}-\ref{i2}) it can be proved that,

\begin{eqnarray} 
\ast \xi_{\alpha\sigma}\:V_{(3)}^{\alpha}&=&
\sqrt{-Q/2}\:V_{(4)\sigma}\label{OR3}\\
\ast \xi_{\alpha\sigma}\:V_{(4)}^{\alpha}&=&
-\sqrt{-Q/2}\:V_{(3)\sigma}\ .\label{OR4}
\end{eqnarray}

If we contract equation (\ref{TWNEXP}) with $\ast \xi_{\alpha\sigma}$ and
make use of identities (\ref{i1}-\ref{i2}) and equations 
(\ref{OR3}-\ref{OR4}) we get,

\begin{equation}
-\sqrt{-Q/2}\:\ast \xi_{\sigma\lambda}\:\ast \Lambda^{\lambda}=
M\:V_{(4)\sigma} - N\:V_{(3)\sigma}\ .\label{BPS}
\end{equation}

This last equation means that,

\begin{equation}
\tilde{V}_{(3)}^{\alpha} = V_{(3)}^{\alpha} + 
N\:V_{(3)}^{\alpha} - M\:V_{(4)}^{\alpha}\ .\label{TZB}
\end{equation}

Then, from (\ref{TZN}) and (\ref{TZB}) we have the following 
relations between coefficients, 

\begin{eqnarray}
K &=& N \\
L &=& -M \ .
\end{eqnarray}

Contracting equation (\ref{BPS}) with $V_{(3)}^{\sigma}$ and
$V_{(4)}^{\sigma}$ it can be found that,

\begin{eqnarray} 
M&=&(-Q/2)\:V_{(3)\sigma}\:\ast \Lambda^{\sigma} / (\:V_{(4)\beta}\:
V_{(4)}^{\beta}\:)\label{COEFFM}\\
N&=&(-Q/2)\:V_{(4)\sigma}\:\ast \Lambda^{\sigma} / (\:V_{(3)\beta}\:
V_{(3)}^{\beta}\:)\ .\label{COEFFN}
\end{eqnarray}

We would like to calculate the norm of the transformed vectors
$\tilde{V}_{(3)}^{\alpha}$ and $\tilde{V}_{(4)}^{\alpha}$,

\begin{eqnarray}
\tilde{V}_{(3)}^{\alpha}\:\tilde{V}_{(3)\alpha} &=& 
[(1+N)^2+M^2]\:V_{(3)}^{\alpha}\:V_{(3)\alpha}\label{FPS}\\
\tilde{V}_{(4)}^{\alpha}\:\tilde{V}_{(4)\alpha} &=& 
[(1+N)^2+M^2]\:V_{(4)}^{\alpha}\:V_{(4)\alpha}\ ,\label{SPS}
\end{eqnarray}
 
where the relation $V_{(3)}^{\alpha}\:V_{(3)\alpha}=
V_{(4)}^{\alpha}\:V_{(4)\alpha}$ has been used.
Now, we see that the gauge transformations
of the invariants given by (\ref{FPS}-\ref{SPS}) cannot change the
spacelike character of vectors $V_{(3)}^{\alpha}$ and $V_{(4)}^{\alpha}$,
unless $1+N=M=0$.
Apart from that exception, the factor $[(1+N)^2+M^2]$ 
is always positive, so we would have no problems 
normalizing the transformed vectors
$\tilde{V}_{(3)}^{\alpha}$ and $\tilde{V}_{(4)}^{\alpha}$,

\begin{eqnarray}
{\tilde{V}_{(3)}^{\alpha} 
\over \sqrt{\tilde{V}_{(3)}^{\beta}\:\tilde{V}_{(3)\beta}}}&=& 
{(1+N) \over \sqrt{(1+N)^2+M^2}}
\:{V_{(3)}^{\alpha} \over \sqrt{V_{(3)}^{\beta}\:V_{(3)\beta}}} -
{M \over \sqrt{(1+N)^2+M^2}}
\:{V_{(4)}^{\alpha} \over \sqrt{V_{(4)}^{\beta}\:V_{(4)\beta}}}\label{TN3}\\
{\tilde{V}_{(4)}^{\alpha} 
\over \sqrt{\tilde{V}_{(4)}^{\beta}\:\tilde{V}_{(4)\beta}}}&=& 
{M \over \sqrt{(1+N)^2+M^2}}
\:{V_{(3)}^{\alpha} \over \sqrt{V_{(3)}^{\beta}\:V_{(3)\beta}}} +
{(1+N) \over \sqrt{(1+N)^2+M^2}}
\:{V_{(4)}^{\alpha} \over \sqrt{V_{(4)}^{\beta}\:V_{(4)\beta}}}\ . 
\label{TN4}
\end{eqnarray}

As long as $[(1+N)^2+M^2]>0$ the transformations 
(\ref{TN3}-\ref{TN4}) are telling us that an
electromagnetic gauge transformation on the vector field $\ast A^{\alpha}$
that leaves invariant the dual electromagnetic field 
$\ast f_{\mu\nu}$, generates
a rotation on the normalized tetrad vector fields 
$\left({V_{(3)}^{\alpha} \over \sqrt{V_{(3)}^{\beta}\:V_{(3)\beta}}},
{V_{(4)}^{\alpha} \over \sqrt{V_{(4)}^{\beta}\:V_{(4)\beta}}}\right)$. 

\section{Group Isomorphism}
\label{Groupiso}

In sections \ref{gaugegeometry1} and \ref{gaugegeometry2} the gauge transformation properties of the tetrad vectors were analyzed. But we can take advantage of the expressions we found for both the transformations on blade one and blade two to prove an important result involving the mappings between the local gauge group and the local Lorentz transformations on both blades. We are going to name LB1 the group of Lorentz transformations on blade one. Analogously we name the group of rotations on blade two, LB2. Making use of expressions (\ref{TN1}-\ref{TN2}), (\ref{TN1SC}-\ref{TN2SC}), (\ref{TN1I}-\ref{TN2I}), (\ref{TN1SCI}-\ref{TN2SCI}) and (\ref{TN3}-\ref{TN4}), we can prove the transformation properties of the local gauge group. We proceed first, to study the transformation properties of the elements of the local gauge group on blade one.

\subsection{Isomorphism on blade one}
\label{Isomor1}

We can readily verify that the identity is given by $\Lambda = 0$ or any other constant. Since all the gauge transformations only involve gradients of scalar functions, then, the results we are going to find are going to be true, except for additional constants, that have no physical or geometrical meaning. If the scalar function $\Lambda$ generates a Lorentz transformation, then ($-\Lambda$) generates the inverse Lorentz transformation. It is necessary at this point to understand several details about inverse transformations. Following the notation of the preceding sections we can introduce on blade one, for instance, the inverse of the direct transformation as,

\begin{eqnarray}
\tilde{\tilde{V}}_{(1)}^{\alpha} &=& \tilde{V}_{(1)}^{\alpha} - 
\xi^{\alpha\lambda}\:\xi_{\rho\lambda}\:\Lambda^{\rho} = V_{(1)}^{\alpha}\label{ITU}\\
\tilde{\tilde{V}}_{(2)}^{\alpha} &=& \tilde{V}_{(2)}^{\alpha} - 
\sqrt{-Q/2}\:\xi^{\alpha\lambda}\:\Lambda_{\lambda} = V_{(2)}^{\alpha}\ ,\label{ITV}
\end{eqnarray}

\begin{eqnarray} 
\tilde{C} &=& (-Q/2)\:\tilde{V}_{(1)\sigma}\:(-\Lambda^{\sigma}) / (\:\tilde{V}_{(2)\beta}\:\tilde{V}_{(2)}^{\beta}\:)\label{ICOEFFC}\\
\tilde{D}&=&(-Q/2)\:\tilde{V}_{(2)\sigma}\:(-\Lambda^{\sigma}) / (\:\tilde{V}_{(1)\beta}\:\tilde{V}_{(1)}^{\beta}\:)\ .\label{ICOEFFD}
\end{eqnarray}

We can see that in expressions (\ref{ICOEFFC}-\ref{ICOEFFD}) the change in sign of the scalar function $\Lambda$ is not the only change. Therefore, we can also notice that for transformations (\ref{TN1}-\ref{TN2}),

\begin{eqnarray} 
{(1+C) \over \sqrt{(1+C)^2-D^2}} =
{(1+\tilde{C}) \over \sqrt{(1+\tilde{C})^2-\tilde{D}^2}} \label{Icoshphi}\\
{D \over \sqrt{(1+C)^2-D^2}} =
-{\tilde{D} \over \sqrt{(1+\tilde{C})^2-\tilde{D}^2}}\ , \label{Isinhphi}
\end{eqnarray}

while for transformations (\ref{TN1I}-\ref{TN2I}),

\begin{eqnarray} 
{(1+C) \over \sqrt{-(1+C)^2+D^2}} = 
-{(1+\tilde{C}) \over \sqrt{-(1+\tilde{C})^2+\tilde{D}^2}} \label{IcoshphiI}\\
{D \over \sqrt{-(1+C)^2+D^2}} =
{\tilde{D} \over \sqrt{-(1+\tilde{C})^2+\tilde{D}^2}}\ . \label{IsinhphiI}
\end{eqnarray}

In order to get a better geometrical insight into the results that follow, we are introducing the angle $\phi$. For proper transformations on blade one,

\begin{eqnarray}
\cosh\phi &=& {\mid 1+C \mid \over \sqrt{(1+C)^2-D^2}}\ , \label{pcoshphi}
\end{eqnarray}

while for improper transformations,

\begin{eqnarray}
\cosh\phi &=& {\mid D \mid \over \sqrt{-(1+C)^2+D^2}}\ . \label{icoshphi}
\end{eqnarray}

For instance, if the scalar function $\Lambda_{1}$ generates a boost $\phi_{1}$, and the scalar function $\Lambda_{2}$ generates a boost $\phi_{2}$, then it is straightforward to see that the subsequent transformation, first by $\Lambda_{1}$ and then by $\Lambda_{2}$ generates a boost $\phi_{1}+\phi_{2}$.
In general, if the scalar function $\Lambda_{1}$ generates a Lorentz transformation on blade one, and the scalar function $\Lambda_{2}$ generates another Lorentz transformation on blade one, then it is straightforward to see that the subsequent transformation, first by $\Lambda_{1}$ and then by $\Lambda_{2}$, generates the composition of the two Lorentz transformations.

Therefore, we proved that the transformations (\ref{TN1}-\ref{TN2}) ,(\ref{TN1SC}-\ref{TN2SC}), (\ref{TN1I}-\ref{TN2I}), (\ref{TN1SCI}-\ref{TN2SCI}), represent a mapping between the local gauge group, and the group LB1. Both groups are Abelian and have the same dimension. If in addition we manage to prove that this mapping is injective, and the image is not a subgroup of LB1, then the mapping would be an isomorphism. To this end, using equations (\ref{COEFFC}-\ref{COEFFD}) and (\ref{COEFFM}-\ref{COEFFN}) we can write,

\begin{eqnarray}
(-Q/2)\:\Lambda^{\alpha} = -C \: V_{(1)}^{\alpha} - D \: V_{(2)}^{\alpha} 
+ M \: V_{(3)}^{\alpha} + N \: V_{(4)}^{\alpha} \ , \label{Lambdainv}  
\end{eqnarray}

such that,

\begin{eqnarray}
D &=& (1+C)\: \tanh\phi \;\;\;\mbox{for proper transformations}\label{DP}\\
D &=& (1+C) / \tanh\phi \;\;\;\mbox{for improper transformations}\ . \label{DI}
\end{eqnarray}

It is simple to check that equation (\ref{DP}) is valid for $1+C > 0$ or 
$1+C < 0$, and equation (\ref{DI}) is also valid for $D > 0$ or $D < 0$.
Once we are given a $\phi$, the functions $C$, $M$ and $N$ should be found through the use of the integrability conditions $\Lambda_{\alpha;\beta} = \Lambda_{\beta;\alpha} $.
We know that if $\phi_{1} \neq \phi_{2}$, then 
$\tanh\phi_{1} \neq \tanh\phi_{2}$ and $\tanh\phi_{1} \neq 1 / \tanh\phi_{2}$. Then, accordingly, the corresponding scalar functions $\Lambda_{1}$ and $\Lambda_{2}$, are not going to be the same. Conversely, we can ask if it is possible to map two different scalar functions $\Lambda_{1}$ and $\Lambda_{2}$ into the same Lorentz transformation. If this is possible, then, we can first generate a Lorentz transformation by $\Lambda_{1}$ and then another one by $-\Lambda_{2}$. The result should be the identity, because $-\Lambda_{2}$ generates the inverse Lorentz transformation of $\Lambda_{1}$. Therefore $\Lambda_{1}-\Lambda_{2}$ must be a constant. Summarizing, the injectivity remains proved.
The last point to make clear is related to the image of this mapping. The question to answer is if the image of this mapping is a subgroup of LB1.
Let us suppose that there is a certain local gauge transformation $\Lambda$, such that $1+C > 0$, with $-1 < C < 0$. Then, there is always the transformation
$n \: \Lambda$ with $n$ a natural number, $C_{n} = n\:C$. $1+C_{n} = 1+n\:C = 1 - n\: \mid C \mid$. For $n$ sufficiently large $1+C_{n}$ is going to become negative. 
If $C > 0$, there is the transformation $-n \: \Lambda$ with $n$ a natural number, $C_{n} = -n\:C$. $1+C_{n} = 1-n\:C = 1 - n\: \mid C \mid$. Once more for $n$ sufficiently large $1+C_{n}$ is going to become negative. Following similar ideas, but now if $1+C > D > 0$ we can prove for instance, that if $-1 < C < 0$, for $n$ sufficiently large, $D_{n} > 0 > 1+C_{n}$, and analogous for $C > 0$. Then, the mapping image, obviously cannot be a subgroup of LB1. Therefore the mapping is surjective.

\newtheorem {guesslb1} {Theorem}
\newtheorem {guesslb2}[guesslb1] {Theorem}
\begin{guesslb1}
The mapping between the local gauge group of transformations is isomorphic, to the group LB1 defined above. 
\end{guesslb1}

\subsection{Isomorphism on blade two}
\label{Isomor2}

The proof for the mapping between the local gauge group and the rotations on blade two is analogous to the previous one. For rotations all the considerations about inverse transformations and composition of transformations apply in a similar fashion as for the ones on blade one. The precisions we have to make on blade two, regard fundamentally the injectivity. We can introduce the angle $\varphi$ as,

\begin{eqnarray}
\cos\varphi &=& {(1+N) \over \sqrt{(1+N)^2+M^2}}\ . \label{cosphi}
\end{eqnarray}

Then, we can use again equation (\ref{Lambdainv}), along with, 

\begin{eqnarray}
M &=& (1+N) \: \tan\varphi \ . \label{M}
\end{eqnarray}

Once we are given a $\varphi$, the functions $C$, $D$ and $N$ should be found through the use of the integrability conditions $\Lambda_{\alpha;\beta} = \Lambda_{\beta;\alpha} $.
For rotations, we know that $\tan(\varphi) = \tan(\varphi-\pi)$, but simultaneously we have that $sg(M_{\varphi}) = - sg(M_{(\varphi-\pi)})$ and 
$sg(1+N_{\varphi}) = - sg(1+N_{(\varphi-\pi)})$. The last sign equalities arise from the fact that under a change $\varphi \rightarrow \varphi-\pi$, the sine and cosine change their signs.
All these results put together mean that given a pair of different angles $\varphi_{1}$ and  $\varphi_{2}$,  the corresponding scalar functions $\Lambda_{1}$ and $\Lambda_{2}$, are not going to be the same. Conversely, we can ask if it is possible to transform two different scalar functions $\Lambda_{1}$ and $\Lambda_{2}$ into the same $\varphi$. If this is possible, then, we can proceed exactly as did for the blade one case, we can first generate a rotation transformation by $\Lambda_{1}$ and then another one by $-\Lambda_{2}$. The result should be the identity, because $-\Lambda_{2}$ generates a rotation transformation by $-\varphi$. Therefore $\Lambda_{1}-\Lambda_{2}$ must be a constant, and the injectivity remains proved. 

\begin{guesslb2}
The mapping between the local gauge group of transformations is isomorphic, to the group LB2 defined above. 
\end{guesslb2}

\section{Tetrads}
\label{tetrads}
\subsection{Orthonormal tetrad}

It was found in section \ref{potentials}, that for an 
electromagnetic vector potential $A^{\alpha}$, 
with $A^{0}\:A_{0} + A^{1}\:A_{1} < 0$ 
in the Minkowski reference frame given by equations (38-39)
in \cite{MW}, it was possible to
normalize the tetrad vectors fields (\ref{V1A}-\ref{V4A}).
Then, at the points in spacetime where the set of four vectors 
(\ref{V1A}-\ref{V4A}) is not trivial, we can proceed to normalize, 

\begin{eqnarray}
U^{\alpha} &=& \xi^{\alpha\lambda}\:\xi_{\rho\lambda}\:A^{\rho} \:
/ \: (\: \sqrt{-Q/2} \: \sqrt{A_{\mu} \ \xi^{\mu\sigma} \ 
\xi_{\nu\sigma} \ A^{\nu}}\:) \label{U}\\
V^{\alpha} &=& \xi^{\alpha\lambda}\:A_{\lambda} \:
/ \: (\:\sqrt{A_{\mu} \ \xi^{\mu\sigma} \ 
\xi_{\nu\sigma} \ A^{\nu}}\:) \label{V}\\
Z^{\alpha} &=& \ast \xi^{\alpha\lambda} \: \ast A_{\lambda} \:
/ \: (\:\sqrt{\ast A_{\mu}  \ast \xi^{\mu\sigma}  
\ast \xi_{\nu\sigma}  \ast A^{\nu}}\:)
\label{Z}\\
W^{\alpha} &=& \ast \xi^{\alpha\lambda}\: \ast \xi_{\rho\lambda}
\:\ast A^{\rho} \: / \: (\:\sqrt{-Q/2} \: \sqrt{\ast A_{\mu} 
\ast \xi^{\mu\sigma} \ast \xi_{\nu\sigma} \ast A^{\nu}}\:) \ .
\label{W}
\end{eqnarray}

The notation we are using to name the four tetrad vectors
(\ref{U}-\ref{W}) is the same
notation used in \cite{HS}, even though the geometrical meaning is different.
The four vectors (\ref{U}-\ref{W}) have the following algebraic properties,

\begin{equation}
-U^{\alpha}\:U_{\alpha}=V^{\alpha}\:V_{\alpha}
=Z^{\alpha}\:Z_{\alpha}=W^{\alpha}\:W_{\alpha}=1 \ .\label{TSPAUX}
\end{equation}

Any other scalar product is zero. 
It is possible to find expressions for the metric tensor and the
stress-energy tensor in the new tetrad 
(\ref{U}-\ref{W}). The new expression for the metric tensor is,

\begin{equation}
g_{\alpha\beta} = -U_{\alpha}\:U_{\beta} + V_{\alpha}\:V_{\beta} +
Z_{\alpha}\:Z_{\beta} + W_{\alpha}\:W_{\beta}\ .\label{MT}
\end{equation}

The stress-energy tensor can be written,

\begin{equation}
T_{\alpha\beta} = (Q/2)\: \left[-U_{\alpha}\:U_{\beta} + 
V_{\alpha}\:V_{\beta} -
Z_{\alpha}\:Z_{\beta} - W_{\alpha}\:W_{\beta}\right]\ .\label{SET}
\end{equation}

In order to find the expression for the electromagnetic field in
terms of the tetrad (\ref{U}-\ref{W}), it is necessary to find some
previous results. Using equations (\ref{i1}) and (\ref{i2}) it is
possible to prove that,

\begin{eqnarray}
U^{\alpha}\:\xi_{\alpha\beta} &=& \sqrt{-Q/2}\:\:V_{\beta}\label{EEF}\\
V^{\alpha}\:\xi_{\alpha\beta} &=& \sqrt{-Q/2}\:\:U_{\beta}\\
Z^{\alpha}\:\ast \xi_{\alpha\beta} &=& \sqrt{-Q/2}\:\:W_{\beta}\\
W^{\alpha}\:\ast \xi_{\alpha\beta} &=& -\sqrt{-Q/2}\:\:Z_{\beta}\ .\label{EE}
\end{eqnarray}

Equations (\ref{TSPAUX}) and (\ref{EEF}-\ref{EE}) allow us to find the 
expressions for the extremal field in terms of the new tetrad,

\begin{eqnarray}
U^{\alpha}\:\xi_{\alpha\beta}\:V^{\beta} &=& \sqrt{-Q/2}\label{ETELF}\\
Z^{\alpha}\:\ast \xi_{\alpha\beta}\:W^{\beta} &=& \sqrt{-Q/2} \ .\label{ETEL}
\end{eqnarray}

The extremal field tensor and its dual can then be written,

\begin{eqnarray}
\xi_{\alpha\beta} &=& -2\:\sqrt{-Q/2}\:U_{[\alpha}\:V_{\beta]}\label{ET}\\
\ast \xi_{\alpha\beta} &=& 2\:\sqrt{-Q/2}\:Z_{[\alpha}\:W_{\beta]}\ .\label{DET}
\end{eqnarray}

Equations (\ref{ET}-\ref{DET}) are providing the necessary information to
express the electromagnetic field in terms of the new tetrad,

\begin{equation}
f_{\alpha\beta} = -2\:\sqrt{-Q/2}\:\:\cos\alpha\:\:U_{[\alpha}\:V_{\beta]} +
2\:\sqrt{-Q/2}\:\:\sin\alpha\:\:Z_{[\alpha}\:W_{\beta]}\ .\label{EMT}
\end{equation}

\subsection{Null tetrad}

We are ready now to introduce a new null tetrad. It will be defined as,

\begin{eqnarray}
K_{\alpha} &=& {1 \over \sqrt{2}}\:\left(U_{\alpha} +
V_{\alpha}\right)\label{K}\\
L_{\alpha} &=& {1 \over \sqrt{2}}\:\left(U_{\alpha} -
V_{\alpha}\right)\label{L}\\
T_{\alpha} &=& {1 \over \sqrt{2}}\:\left(Z_{\alpha} + \imath
W_{\alpha}\right)\label{T}\\
\overline{T}_{\alpha} &=& {1 \over \sqrt{2}}\:\left(Z_{\alpha} - \imath
W_{\alpha}\right)\ .\label{THAT}
\end{eqnarray} 

Where $\imath$ is the imaginary complex unit, $\imath^2=-1$.
The notation we are using to name the null tetrad vectors 
(\ref{K}-\ref{THAT}), is the same as in \cite{HS}.
Again it should be noticed that even though the null tetrad notation
is the same as in \cite{HS}, the geometrical meaning is not the same.
This new null tetrad (\ref{K}-\ref{THAT}) satisfies the 
following algebraic relations,

\begin{equation}
K^{\alpha}\:K_{\alpha}=K^{\alpha}\:T_{\alpha}
=L^{\alpha}\:L_{\alpha}=L^{\alpha}\:T_{\alpha}=T^{\alpha}\:T_{\alpha}=0 \ ,
\label{ANN}
\end{equation}

and,

\begin{eqnarray}
K^{\alpha}\:L_{\alpha} &=& -1\\
T^{\alpha}\:\overline{T}_{\alpha} &=& 1\ .\label{PONN}
\end{eqnarray}

It can also be proved in the Minkowskian reference frame 
given by equations (38-39) in \cite{MW}, and through the use of 
the results found in sections \ref{sec:appI} and \ref{sec:appII} 
that the identity,

\begin{equation}
\epsilon_{\alpha\beta\mu\nu}\:K^{\alpha}\:L^{\beta}\:
T^{\alpha}\:\overline{T}^{\nu} = -\imath\ ,\label{EPID}
\end{equation}

is satisfied. Then, the metric tensor can be written in terms of the 
new null tetrad,

\begin{equation}
g_{\alpha\beta} = T_{\alpha}\:\overline{T}_{\beta} + 
\overline{T}_{\alpha}\:T_{\beta} -
K_{\alpha}\:L_{\beta} - K_{\beta}\:L_{\alpha}\ .\label{MNT}
\end{equation}

Using (\ref{K}-\ref{THAT}) it is straightforward to prove that (\ref{MNT})
is equivalent to (\ref{MT}). In terms of the new null tetrad the 
stress-energy tensor can be expressed,

\begin{equation}
T_{\alpha\beta} = (-Q/2)\: \left[K_{\alpha}\:L_{\beta} + 
K_{\beta}\:L_{\alpha} +
T_{\alpha}\:\overline{T}_{\beta} + 
T_{\beta}\:\overline{T}_{\alpha}\right]\ .\label{SENT}
\end{equation}

It is not difficult to prove that (\ref{SENT}) and (\ref{SET}) are equivalent.
We would like to find the expression for the electromagnetic field
in terms of the new null tetrad. It is necessary to find first the 
components of the extremal tensor and its dual in terms of 
the new null tetrad. Making use of (\ref{ETELF}-\ref{ETEL}) we find,

\begin{eqnarray}
K^{\alpha}\:\xi_{\alpha\beta}\:L^{\beta} &=& -\sqrt{-Q/2}\\
T^{\alpha}\:\ast \xi_{\alpha\beta}\:\overline{T}^{\beta} 
&=& -\imath\:\sqrt{-Q/2} \ .\label{EENT}
\end{eqnarray}

Now, we have the expressions for the extremal field tensor and its dual 
in terms of the new null tetrad,

\begin{eqnarray}
\xi_{\alpha\beta} &=& 2\:\sqrt{-Q/2}\:K_{[\alpha}\:L_{\beta]}\label{EKL}\\
\ast \xi_{\alpha\beta} &=& 2\:\imath\:\sqrt{-Q/2}\:
T_{[\alpha}\:\overline{T}_{\beta]}\ .\label{DETTHAT}
\end{eqnarray}

Equations (\ref{EKL}-\ref{DETTHAT}) allow us to find the electromagnetic 
field expressed in terms of the new null tetrad,

\begin{equation}
f_{\alpha\beta} = 2\:\sqrt{-Q/2}\:\:\cos\alpha\:\:K_{[\alpha}\:L_{\beta]} +
2\:\imath\:\sqrt{-Q/2}\:\:\sin\alpha\:\:
T_{[\alpha}\:\overline{T}_{\beta]}\ .\label{EMNT}
\end{equation}

\subsection{Bivectors}

It is possible to express any antisymmetric second-rank tensor,
as a linear combination of the following bivectors,

\begin{eqnarray}
U_{\alpha\beta} &=& \overline{T}_{\alpha}\:L_{\beta} -
\overline{T}_{\beta}\:L_{\alpha}\label{UB}\\
V_{\alpha\beta} &=& K_{\alpha}\:T_{\beta} -
K_{\beta}\:T_{\alpha}\label{VB}\\
W_{\alpha\beta} &=& T_{\alpha}\:\overline{T}_{\beta} -
T_{\beta}\:\overline{T}_{\alpha} -
K_{\alpha}\:L_{\beta} + K_{\beta}\:L_{\alpha}\ .\label{WB}
\end{eqnarray}

These bivectors are combinations of the null tetrad vectors 
(\ref{K}-\ref{THAT}), and making use of (\ref{EPID}) it can be
proved that they are also self-dual,

\begin{eqnarray}
\tilde{U}_{\alpha\beta} &=& {1 \over 2}\: \epsilon_{\alpha\beta\mu\nu}\:
U^{\mu\nu} = -\imath\:U_{\alpha\beta}\\
\tilde{V}_{\alpha\beta} &=& -\imath\:V_{\alpha\beta}\\
\tilde{W}_{\alpha\beta} &=& -\imath\:W_{\alpha\beta}\ .\label{SD}
\end{eqnarray}

One more time, the notation for (\ref{UB}-\ref{WB}) is analogous to the
one in \cite{HS} but the geometrical meaning is different.
The self-dual bivectors (\ref{UB}-\ref{WB}) have the following
associated scalar products, 

\begin{equation}
W_{\alpha\beta}\:V^{\alpha\beta} = W_{\alpha\beta}\:U^{\alpha\beta} =
V_{\alpha\beta}\:V^{\alpha\beta} = U_{\alpha\beta}\:U^{\alpha\beta} = 0\ ,
\label{BSP1}
\end{equation}

\begin{eqnarray}
W_{\alpha\beta}\:W^{\alpha\beta} &=& -4\\
U_{\alpha\beta}\:V^{\alpha\beta} &=& 2\ .\label{BSP2}
\end{eqnarray}

It was our purpose since the beginning of this work,
to find the simplest possible expression for the electromagnetic field,
through the use of null tetrads in geometrodynamics.
Now, we would like to see what is the expression in the new bivectors 
(\ref{UB}-\ref{WB}) for the self-dual electromagnetic bivector,

\begin{equation}
\Phi_{\alpha\beta} = f_{\alpha\beta} + 
\imath \ast f_{\alpha\beta}\ .\label{SDB}
\end{equation}
 
Making use of expressions (\ref{EKL}-\ref{DETTHAT}) and (\ref{WB}) 
it is possible to write,

\begin{equation}
\Phi_{\alpha\beta} = -\sqrt{-Q/2}\:e^{(-\imath\:\alpha)}\:W_{\alpha\beta}\ .
\label{SDEMNT}
\end{equation}

The standard way of expressing the bivector (\ref{SDB}) is through
an expansion in the three standard bivectors given by equation (18.9)
in \cite{HS}, which in turn are built in terms of the familiar 
NP tetrads \cite{NP}.
The new null tetrad (\ref{K}-\ref{THAT}) has the advantage that in the 
particular problem of geometrodynamics the expression of the 
bivector (\ref{SDB}) given by (\ref{SDEMNT})
is the simplest possible one, since we can write it only in terms
of one of the three independent new bivectors, expression (\ref{WB}). 
The two scalar
functions associated with the electromagnetic field are included in the
complex factor $-\sqrt{-Q/2}\:e^{(-\imath\:\alpha)}$. This example shows
the simplifying power of the new tetrad built specifically for the problem
of geometrodynamics. 

\section{General tetrad}
\label{Gentetrads}

In section (\ref{potentials}) a particular example of an explicit
choice of the vector fields $X_{\alpha}$ and $Y_{\alpha}$ was introduced
($X_{\alpha}=A_{\alpha}$ and $Y_{\alpha}=\ast A_{\alpha}$)
within the framework of geometrodynamics. Through this example several
properties of these new tetrads were discussed. The transformation
properties induced by usual gauge transformations 
$A^{\alpha} \rightarrow A^{\alpha} + \Lambda^{,\alpha}$ and 
$\ast A^{\alpha} \rightarrow \ast A^{\alpha} + \ast \Lambda^{,\alpha}$
were analyzed in section (\ref{gaugegeometry}).
Let's assume that it is possible to normalize the general tetrad
(\ref{V1}-\ref{V4}). For generic fields $X_{\alpha}$ and $Y_{\alpha}$, 
we can proceed to study the transformation
properties of the normalized version of (\ref{V1}-\ref{V2}),
under the transformation $X_{\alpha} \rightarrow X_{\alpha} + V_{\alpha}$,
where $V_{\alpha}$ is any well-behaved vector field in spacetime.
The necessary steps to study these tetrad transformations are a replica
of the ones taken in section (\ref{gaugegeometry1}). The conclusions
are analogous. In a similar way, the normalized
version of (\ref{V3}-\ref{V4}) can be transformed under 
$Y_{\alpha} \rightarrow Y_{\alpha} + W_{\alpha}$, where $W_{\alpha}$ is
any well-behaved vector field. Again, the steps and conclusions involved in
the study of these tetrad transformations are a replica
of the ones taken in section (\ref{gaugegeometry2}).
Once we introduced these new tetrad transformations for the general case, 
we can easily
prove that $X_{\alpha} \rightarrow X_{\alpha} + V_{\alpha}$ and
$Y_{\alpha} \rightarrow Y_{\alpha} + W_{\alpha}$ leave invariant two
tensors, $g_{\alpha\beta}$ and $f_{\alpha\beta}$.
The fact that these tetrad transformations leave invariant the metric
tensor means that they are symmetries of the geometry. They also leave invariant
the electromagnetic tensor, which means that they represent 
symmetries of the electromagnetic geometry. But the important issue
is that they do not alter the tensors that carry the physical information.
The new tetrads (\ref{V1}-\ref{V4}) when normalized can
be used to simplify the Einstein-Maxwell equations in the Newman-Penrose
tetrad formalism. 
It is worth noticing the
substantial simplification that the new tetrads would introduce
in the Einstein-Maxwell equations. As an example, the vacuum Maxwell
equations using the notation given in \cite{MC} are given by

\begin{eqnarray} 
D\phi_{1}&=& 2\,\rho\,\phi_{1}\label{EMNT1}\\
\delta\phi_{1}&=&-2\,\tau\,\phi_{1}\label{EMNT2}\\
\overline{\delta}\phi_{1}&=&-2\,\pi\,\phi_{1}\label{EMNT3}\\
\Delta\phi_{1}&=&-2\,\mu\,\phi_{1} \ .\label{EMNT4}
\end{eqnarray}

It is important to notice that the tetrads will be independent variables.
For a non-null solution to the set of Einstein-Maxwell equations
we can find the scalars $\alpha$ and $Q$ from 
$\phi_{1}=-\sqrt{-Q/2}\:e^{(-\imath\:\alpha)}$.
Expression (\ref{EKL}) will provide the relation between the 
extremal field and the new tetrads.
The electromagnetic field will be available through expression (\ref{EMNT})
and the metric tensor will be given by (\ref{MNT}) once a tetrad
is known for a particular solution. 
The overall effect of the new tetrads is to reallocate or reorganize 
algebraic information in the Einstein-Maxwell equations making it easier
to find new possible solutions.

\section{Conclusions}

A new tetrad that diagonalizes the electromagnetic stress-energy tensor
for non-null electromagnetic fields was introduced. 
However, this tetrad has an inherent freedom in the
choice of two vector fields. Geometrodynamics is an arena in
which an explicit example or choice can be given for these two 
vector fields, because
Maxwell's equations are providing two vector potentials. The simplicity
of the expression for the electromagnetic field in this new tetrad and
the associated null tetrad becomes evident. It was also proved that the local gauge group is related to the group LB1, and also to the group LB2 through an isomorphism.
In the last section the tetrad is also considered without 
making any specific choice
for these two vector fields. It is found that there are transformations
that leave invariant the metric and electromagnetic tensors simultaneously.
It was also proved that when written in terms of the new tetrad, 
the Einstein-Maxwell equations are substantially simplified.     




\section{Appendix I}
\label{sec:appI}

The Levi-Civita pseudotensor can be transformed into a tensor through
the use of factors $\sqrt{-g}$, where $g$ is the determinant of the
metric tensor. We use the notation 
$e_{\alpha\beta\mu\nu}=[\alpha\beta\mu\nu]$ for the 
covariant components of the Levi-Civita pseudotensor in the 
Minkowskian frame given in \cite{MW},

\begin{center}
$ e_{\alpha\beta\mu\nu} = \left\{ \begin{array}{ll}
				1 \:\: \mbox{if $\alpha\beta\mu\nu$ is an even permutation of 0123}\\
				-1 \:\: \mbox{if $\alpha\beta\mu\nu$ is an odd permutation of 0123}\\
				0 \:\: \mbox{if $\alpha\beta\mu\nu$ are not all different}
				    \end{array}
			    \right. $
\end{center}

It can be noticed that the signs in $e^{\alpha\beta\mu\nu}$ are going 
to be opposite to the standard notation \cite{WE}. 
The reason for this is that we want to keep
the compatibility with \cite{MW} where the definition 
$e_{0123}=[0123]=1$ was adopted.
With these definitions we see that in a spacetime 
with a metric $g_{\alpha\beta}$,

\begin{equation}
\epsilon^{\alpha\beta\mu\nu}= 
{e^{\alpha\beta\mu\nu} \over  \sqrt{-g}}= 
- {[\alpha\beta\mu\nu] \over \sqrt{-g}} \ ,\label{lccon}
\end{equation}

are the components of a contravariant tensor \cite{WE}\cite{LL}\cite{MC}. 
The covariant components of (\ref{lccon}) are

\begin{equation}
\epsilon_{\alpha\beta\mu\nu}= e_{\alpha\beta\mu\nu} \sqrt{-g}=
[\alpha\beta\mu\nu] \sqrt{-g} \ ,\label{lccov}
\end{equation}

where

\begin{equation}
g_{\alpha\sigma} g_{\beta\rho} 
g_{\mu\kappa} g_{\nu\lambda}\:e^{\sigma\rho\kappa\lambda}= -g\:
e_{\alpha\beta\mu\nu}\ ,  
\end{equation}

is satisfied.

\section{Appendix II}
\label{sec:appII}

The tetrad vectors $(U^{\alpha},V^{\alpha},Z^{\alpha},W^{\alpha})$
have the following expressions in the Minkowski reference frame given
by equations (38-39) in \cite{MW},

\begin{eqnarray}
U^{0} &=& - A^{0} / \sqrt{-(A^{0}\:A_{0} + A^{1}\:A_{1})}\\
U^{1} &=& - A^{1} / \sqrt{-(A^{0}\:A_{0} + A^{1}\:A_{1})}\\
V^{0} &=& - \xi_{01} \: A^{1} / 
\left(\mid\xi_{01}\mid\:\sqrt{-(A^{0}\:A_{0} + A^{1}\:A_{1})}\right)\\
V^{1} &=& - \xi_{01} \: A^{0} / 
\left(\mid\xi_{01}\mid\:\sqrt{-(A^{0}\:A_{0} + A^{1}\:A_{1})}\right)\\
Z^{2} &=& - \xi_{01} \: \ast A^{3} / 
\left(\mid\xi_{01}\mid\:
\sqrt{\ast A^{2}\:\ast A_{2} + \ast A^{3}\:\ast A_{3}}\right)\\
Z^{3} &=& \xi_{01} \: \ast A^{2} / 
\left(\mid\xi_{01}\mid\:
\sqrt{\ast A^{2}\:\ast A_{2} + \ast A^{3}\:\ast A_{3}}\right)\\
W^{2} &=& \ast A^{2} / 
\sqrt{\ast A^{2}\:\ast A_{2} + \ast A^{3}\:\ast A_{3}}\\
W^{3} &=& \ast A^{3} / 
\sqrt{\ast A^{2}\:\ast A_{2} + \ast A^{3}\:\ast A_{3}}\ ,\label{TMRF}
\end{eqnarray}

where $\mid\xi_{01}\mid\ = \sqrt{(\xi_{01})^2}$.

\acknowledgements 

I am grateful to R. Gambini and J. Pullin for reading this manuscript and
for many fruitful discussions. This work was partially funded by PEDECIBA.

\end{document}